\documentclass[review,number,sort&compress]{elsarticle}
\journal{Nuclear Instruments and Methods in Physics Research A}

\usepackage[english]{babel}
\usepackage[utf8]{inputenc}
\usepackage[T1]{fontenc}
\usepackage{hyperref}
\usepackage{graphicx}
\usepackage{color}
\usepackage{amsmath}

\def\Emax{E_{\mathrm{max}}}
\def\parton#1{\left({{#1}}\right)}

\def\aladyn{\texttt{ALaDyn} }
\begin{document}

\title{TNSA proton maximum energy laws for 2D and 3D PIC simulations}

\author[unibo]{S.~Sinigardi\corref{cor1}}
\ead{sinigardi@bo.infn.it}
\cortext[cor1]{Corresponding author}
\author[iran]{J.~Babaei}
\author[unibo]{G.~Turchetti}

\address[unibo]{Dipartimento di Fisica e Astronomia, Universit\`a di Bologna and INFN Sezione di Bologna, Via Irnerio 46, I-40126 Bologna (BO), Italy}
\address[iran]{Department of Physics, Faculty of Basic Sciences, University of Mazandaran, P. O. Box 47415--416, Babolsar, Iran}

\date{\today}

\begin{abstract}
Numerical simulations are a prominent tool in laser-plasma experiments. Their role, as a guide in new regime explorations and as a support for understanding laboratory results, is undisputed. But as the experiments themselves are growing in costs, setup time and complexity, so are the numerical counterparts. Nowadays it is often necessary to investigate, with great accuracy, a huge set of parameters, in order to explore interesting features.

In literature, it is well known that two-dimensional particle in cell (2D PIC) simulations can only give a qualitative estimation of experimental results, often through a great layer of arbitrariness. On the other hand, three-dimensional (3D) PIC simulations, for the same setup, can typically require two orders of magnitude more of computational resources, to deliver results that, while being similar to laboratory results, are still far from being a real match, due to the many uncertainties, in the parameters and in the model, included in the physical engine.

Following our recent published work, we discuss here a couple of empirical laws that we proposed, that can help giving quantitative insight into 2D PIC simulations and determining when 3D simulations should be stopped, if it is not really necessary to do a detailed exploration of the numerical results at long times.
\end{abstract}

\maketitle

\section{Introduction}\label{intro}
Laser plasma acceleration is evolving at a fast pace on both fronts, experimental and numerical. Especially for ion acceleration, the quest for a stable, reproducible and convenient beam is still far from over.

While laser have now reached the multi-PetaWatt power, also for pulses in the range of tens of femtoseconds (25--50 fs), and targets can be manufactured with extremely small thicknesses and many special layers on the illuminated side, to enhance coupling and energy conversion efficiency, nowadays simulations can run on tens of thousands of CPU cores, each one able to perform up to TFLOPS, exploiting accelerating devices like GPU, consuming tens of terabytes of RAM and storage for each run.

It is very easy to understand why the necessity to optimize simulation codes is now as important as having good optics to transport the laser beam to the target. In an ever increasing moral and environmental journey to be also energy efficient, being able to cut simulation computing requirements can have a long-haul effect.
To reduce simulation times we can proceed on two different fronts: we have to pursue the everlasting quest to improve the code, removing bottlenecks, rewriting slower parts of the software, implementing modern toolchains and updated libraries, or we can completely rethink the problem and tackle it from a different point of view.

To have quantitatively meaningful results, traditionally we cannot rely on 2D PIC simulations, which can give only qualitative insights. But moving to a 3D PIC simulation means increasing our computing requests by a factor of more than \( 100\times \).

In a recent work, we proposed~\cite{Babaei} two empirical laws to describe the rise in time of the cut-off energy, both for 2D and 3D PIC simulations, for the Target Normal Sheath Acceleration (TNSA) regime. This semi-analytic, specially tailored towards numerical better interpretation, can be regarded as one of the methods to drastically cut the computational costs of simulations, being able to reduce them by a factor of up to one order of magnitude in some specific conditions.

\section{Two phenomenological laws}\label{laws}
Our work~\cite{Babaei} produced a couple of phenomenological laws for \( \Emax(t) \), one suggested by a model firstly proposed by Schreiber et al.~\cite{Schreiber} for the 3D case, while the other one is an extension of the aforementioned model for 2D PIC simulations proposed by the authors.

In TNSA, the proton acceleration is due to an electric field generated by charge separation, created by the interaction on a solid target of an extremely intense laser pulse, which pushes out the electrons (mostly on the rear side), that finally extracts protons inside the contaminant layers, due to the huge electric fields induced in the process.

Schreiber et al.~\cite{Schreiber} proposed a simple theoretical approach to the model, describing the acceleration as due to a positive surface charge on the non-illuminated side. Despite its naive description, the model is quite powerful, especially when applied to the rise empirical laws.

For an overview on the physics of the proton acceleration by high intensity lasers and related experiments, we refer to recent reviews~\cite{Borghesi,Macchi_2_RMP,Daido}. In the intensity range that we have considered, experimental results concerning the dependence on the target thickness, the incidence angle and the temporal contrast are reported in many papers~\cite{Fritzler,Ceccotti,Zeil,Spencer,Neely,Yogo,Flacco}.

As we know, the energy spectra in TNSA are found to be exponential with a cut-off (\( \Emax \)). In 2D PIC TNSA simulations, \( \Emax \) grows monotonically during the numerical run; on the other hand, plots for 3D simulations suggest a possible saturation, which is anyway never observed numerically due to the extreme computational costs required to reach such a final configuration.

For these reasons, 2D simulations cannot be used to predict energies: an arbitrary choice of a different simulation end-of-run time would bring different results. To our knowledge, there is no other proposed law in literature to explain how to deal with such a key parameter. Usually, authors rely on experience or some magic values (like twice the laser pulse duration), which doesn't offer very clear explanation of their reliability.

\subsection{3D}
To summarize the results of ref.~\cite{Babaei}, we consider a laser pulse which propagates along the \( z \) axis and choose an electrostatic potential which vanishes at \( z=0 \), where a uniform charge density \( \sigma \), within a disc of radius \( R \), is located. A particle initially at rest accelerates and its law of motion is obtained from energy conservation. The kinetic energy of the particle, obtained integrating the equation of motion, is
\begin{equation}
E(t) \simeq E_\infty\parton{ 1-{t^*\over t}}^2 \qquad \quad t>t^* = {R\over 4v_\infty}
\label{eq:3d}
\end{equation}
where
\[ E_\infty= m{v_\infty^2\over 2}=2\pi eR\sigma \]

\subsection{2D}
In our extension to a 2D geometry, proposed in~\cite{Babaei}, we considered an infinite strip along the \( y \) axis, with a uniform charge density \( \sigma \) on \( -R<x<R \). We approximated the potential energy with
\[ e\hat{V}(z)= - E_\infty \, \log(1+\zeta) \qquad \qquad
E_\infty\equiv m{v_\infty^2\over 2}= 4eR\sigma \]
and again solved the equations of motion from energy conservation, assuming the proton initially at rest at the origin as for the 3D case. The result is
\begin{equation}
E(t)= E_\infty \log\parton{t\over t^*}
\qquad \qquad
t\geq t^*= {R\over v_\infty}
\label{eq:2d}
\end{equation}

\section{Results for simulations}
The kinetic energy of the protons follows two distinct phenomenological laws, depending on the dimensionality of the simulation. Both of these laws appear to be nicely satisfied by PIC simulations, considering a model with a target given by a uniform foil plus a hydrogen-rich layer. The laws, as we can see from the formulas above, depend on two parameters: the scaling time, at which the energy starts to rise, and the asymptotic cut-off energy.

In our model, the preplasma is neglected (the temporal contrast is assumed as infinite).

The 2D and 3D simulations were carried out with the \aladyn{} code~\cite{ALaDyn_GPL_20171123} and the asymptotic cut-off energy \( E_\infty \) was determined by a best-fit procedure on its time dependence, following the laws obtained from the electrostatic models described above.

We tested our laws with a single configuration of a laser pulse impinging on various target thicknesses. More results will come in a following work.

We have considered the following setup: a laser pulse, with wavelength \( \lambda=0.8 \, \mu \)m, intensity \(I=2 \, \cdot \, 10^{19}  \) W/cm\(^2\), waist 6.2 \( \mu \)m, P-polarization and duration is 40 fs, whose corresponding normalized vector potential is \(a_0=3 \), and a target made by a uniform Al foil, having thickness \( L \) varying between 0.5 and 8 \( \mu \)m, having a layer of hydrogen on the rear (non illuminated) side, with fixed thickness \(0.08 \mu \)m.

The ionization level is Al\(^{9+}\) and H\(^{+}\) and it is fixed throughout the simulation. The electron densities have been chosen as \(n_e^{Al}=100 \,n_c \) and \(n_e^H=10\,n_c \).

The law to be fitted for 3D simulations is eq.~\ref{eq:3d}, which can be fitted linearly by defining \(y=\sqrt{E} \) and \(x=1/ct \), so that the previous law becomes
\[ y=a+bx \qquad \qquad \qquad E^{(3D)}_\infty=a^2 \qquad t^{*(3D)}= -{b\over a} \]

The law to be fitted for 2D simulations is eq.~\ref{eq:2d}, which is easily linearly fitted by defining \(y=E \) and \(x=\log ct \), so that the previous law becomes
\[ y=a+bx \qquad \qquad \qquad E^{(2D)}_\infty=b \qquad t^{*(2D)}=e^{-a/b} \]

\begin{figure}
\centering
\includegraphics[width=0.85 \textwidth]{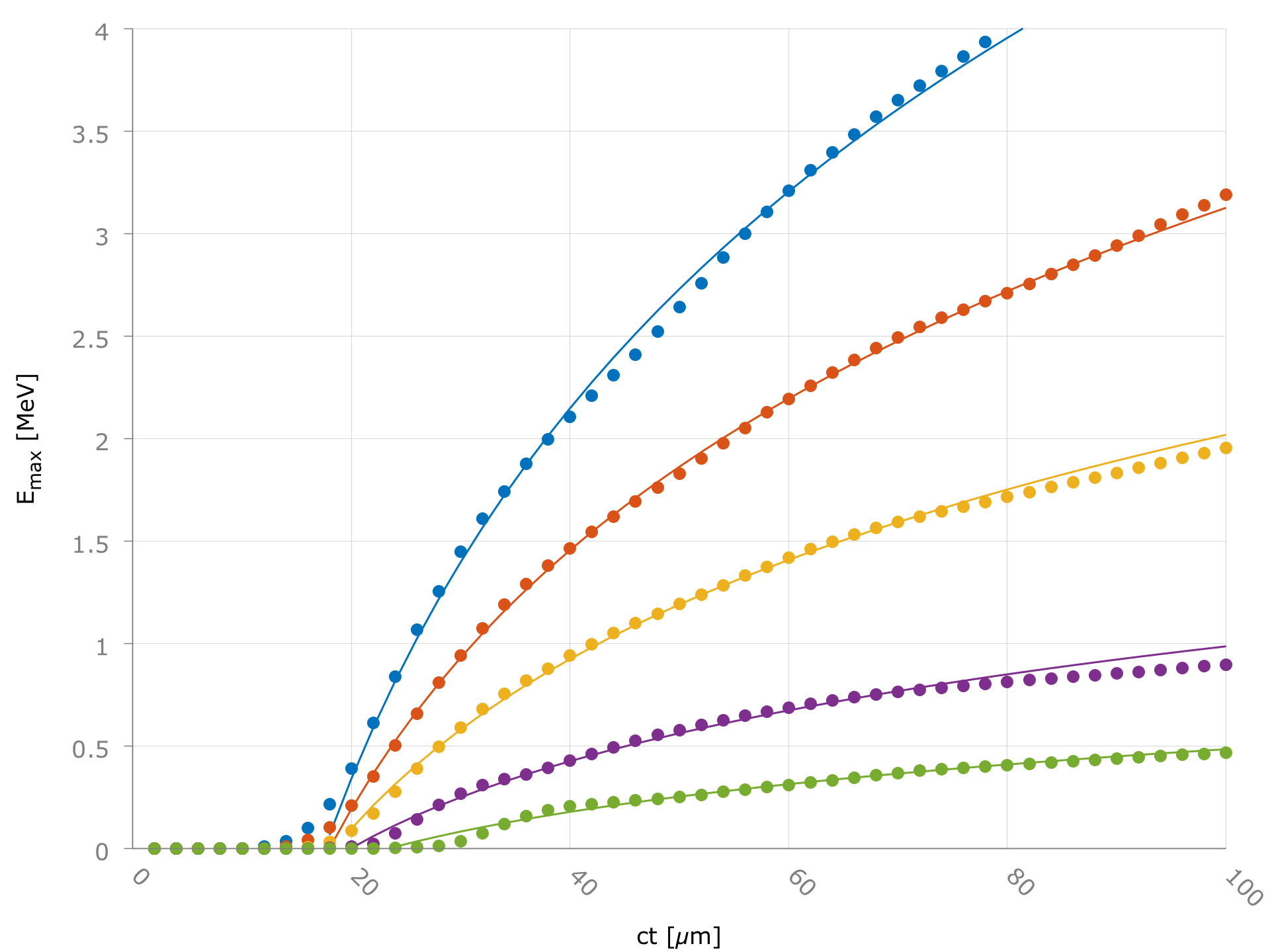}
\caption{Cut-off energy \( \Emax \) versus \(ct \) in the range \(10\leq ct\leq 100 \, \mu \)m obtained from a PIC simulation (points) and comparison with the fit (continuous line) for targets of various thicknesses \(L \): blue \(L=0.5 \, \mu \)m, orange \(L=1 \, \mu \)m, yellow \(L=2 \, \mu \)m, violet \(L=4 \, \mu \)m, green \(L= \,8\mu \)m.}\label{fig:2d_fit}
\end{figure}

\begin{figure}
\centering
\includegraphics[width=0.85 \textwidth]{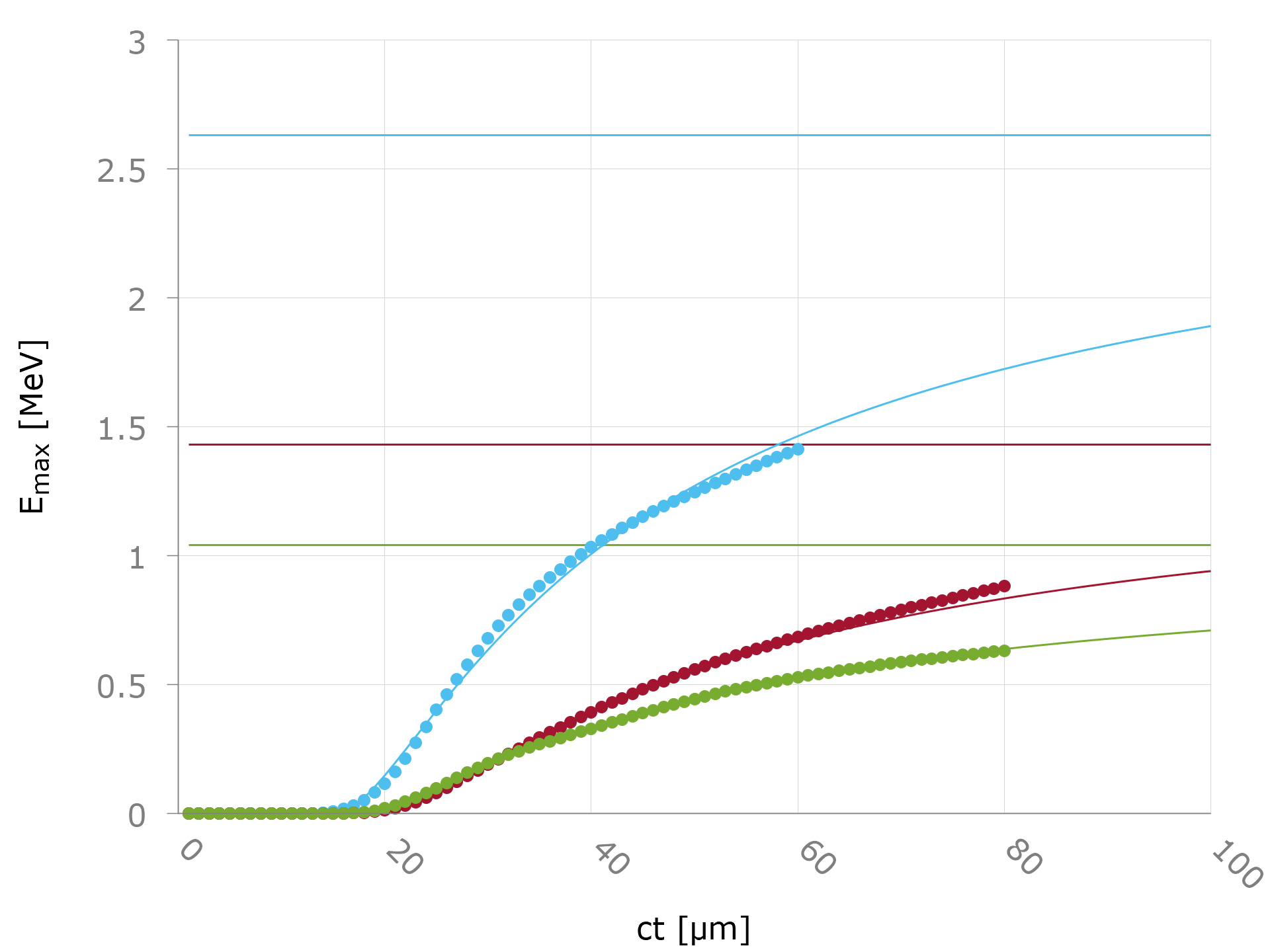}
\caption{Results for a 3D PIC simulation for \( \Emax  \) versus \( ct \) (points) compared with the linear fit of \( \sqrt{\Emax} \) as a function of \(1/ct \) (continuous lines, the asymptotic values \( E_\infty^{(3D)} \) are also shown), for different target thickness: \(L=0.5 \mu \)m cyan, \(L=1 \mu \)m red, \(L=2 \mu \)m green.}\label{fig:3d_fit}
\end{figure}

A plot showing good agreement between the fitting law and 2D simulations can be seen in figure~\ref{fig:2d_fit}, while for 3D simulations we can analyse figure~\ref{fig:3d_fit}.

Our method allows us to limit the simulation even to \(ct= 60 \sim 80\, \mu \)m, which in turn allows using also smaller numerical boxes to contain all the particles. At these times, results are already perfectly fittable.

2D simulations were performed using a grid of \( 6000 \times 1200 \) points, with 100 points per \( \mu \)m and 120 particles per cell. The computational requirements were usually below the 100 core-hours to reach \(ct = 80\, \mu \)m, and can even run on a multi-core single CPU system.

In 3D simulations, as clearly shown in fig.~\ref{fig:3d_fit}, the expected asymptotic limit is reached within \( 5\% \) only quite far, when \(ct > 200 \mu \)m, which is computationally too expensive to be attained. A correct extrapolation, anyway, is still possible with data just up to \(ct \leq 50 \, \mu \)m.

3D simulations were performed using a grid of \( 4096 \times 1536 \times 1536 \) points, with 70 points per \( \mu \)m and \( \sim 16 \) particles per cell. The computational requirements were usually above the 20000 core-hours to reach \(ct = 50\, \mu \)m, and requires an HPC facility with at least 2048 cores dedicated to each single simulation. 

\section{Comparison with experiments}
\begin{figure}
\centering
\includegraphics[width=0.85 \textwidth]{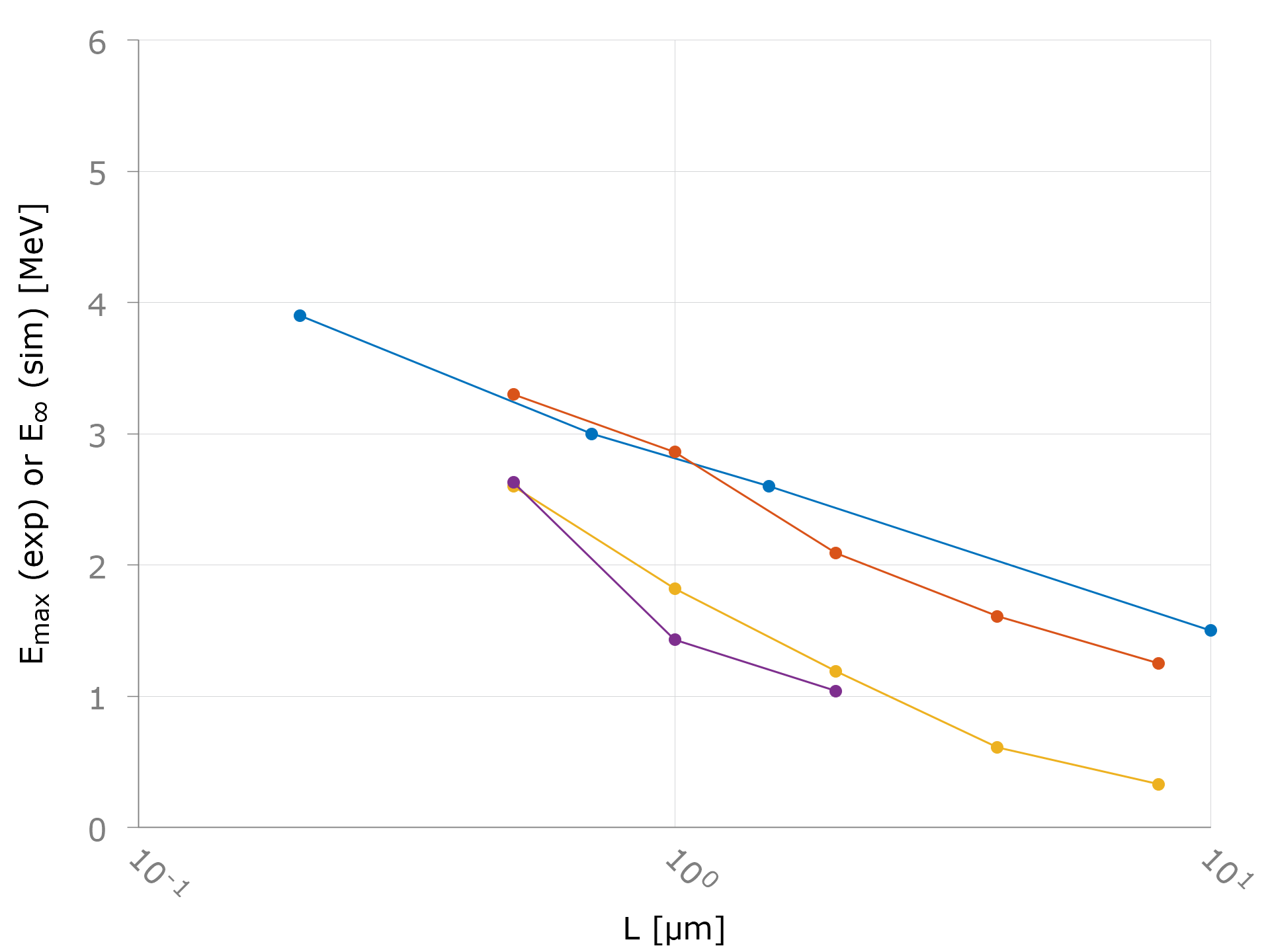}
\caption{Plot of \( \Emax \) versus \( L \) (in logarithmic scale) from an experiment with a laser pulse having \( a_0\sim 3 \) and a metal target, with incidence of 30\(^\circ \) (blue)~\cite{Neely}. These data are compared with \( E_\infty \) obtained fitting (using our laws) 2D PIC simulations of the experiment (red). We also show, on the same plot, a comparison between results of extrapolated maximum energy (\( E_\infty \)) from 2D PIC simulations (yellow) and 3D PIC simulations (violet) using the same setup but with normal incidence. 3D simulation ran up to \(ct = 50\, \mu \)m while the saturation level extrapolated from our laws would have been reached at \(ct > 200 \, \mu \)m}\label{fig:comparison_with_experiments}
\end{figure}

Finally we tested our laws with some experimental results, in particular from one experiment~\cite{Neely} which was close to our interests and whose configuration was similar to the one assumed in our work. Instead of analysing the rise in time of the \( \Emax \) in this case we are just interested in the maximum energy attainable in each setup. We ran some 2D PIC simulations and applied our phenomenological law to extract a proper value for the expected cut-off energy. The results, with a surprisingly good agreement with the experimental data, are shown in figure~\ref{fig:comparison_with_experiments}.

\section{Conclusions}

The asymptotic value of the cut-off energy of protons, which is what is measured in experiments, is difficult to extract from PIC simulations. It is known that 2D results do not exhibit a saturation in \( \Emax \) growth in time, whereas 3D results show that a saturation might be reached, despite requiring a large computational time (\( ct > 200 \mu \)m), which is too expensive to be reached. To ease this problem, we formulated two empirical laws, for 2D and 3D simulations, which depend on the asymptotic energy \( E_\infty \). The fits to the 2D and 3D results coming from PIC simulations are quite good and the statistical uncertainties are a few percent. As discussed in~\cite{Babaei}, the extrapolated values \( E_\infty^{(2D)} \) and \( E_\infty^{(3D)} \) are comparable and can be fully calculated fitting the results obtained before \( ct < 50 \sim 60 \mu \)m, which is a distance reachable also in 3D simulations.

The proposed phenomenological model is adequate to avoid the arbitrariness in the choice of the time at which the asymptotic cut-off energy is chosen in numerical simulations, offering a way to minimize the run time to the least data sufficient for a proper fit. Two-dimensional simulations may offer a quantitative insight, with an adequate extrapolation, rather than being of purely qualitative nature.

\section*{Acknowledgements} The work has been done within the L3IA INFN Collaboration, which the authors would like to thank all. The numerical work has been supported by CINECA through grant ISCRA-B:\@ IsB13\_AGOS.\@


\section*{References}
\begin{thebibliography}{10}

\bibitem{Babaei}
J.~Babaei, L.~A. Gizzi, P.~Londrillo, S.~Mirzanejad, T.~Rovelli, S.~Sinigardi, and G.~Turchetti.
\newblock{} Rise time of proton cut-off energy in 2d and 3d pic simulations.
\newblock{} {\em Physics of Plasmas}, 24(4):043106, 2017.

\bibitem{Schreiber}
J.~Schreiber, F.~Bell, F.~Gr\"uner, U.~Schramm, M.~Geissler, M.~Schn\"urer, S.~Ter-Avetisyan, B.~M. Hegelich, J.~Cobble, E.~Brambrink, J.~Fuchs, P.~Audebert, and D.~Habs.
\newblock{} Analytical model for ion acceleration by high-intensity laser pulses.
\newblock{} {\em Phys. Rev. Lett.}, 97:045005, Jul 2006.

\bibitem{Borghesi}
M.~Borghesi, J.~Fuchs, S.~V. Bulanov, A.~J. Mackinnon, P.~K. Patel, and M.~Roth.
\newblock{} Fast ion generation by high-intensity laser irradiation of solid targets and applications.
\newblock{} {\em Fusion Science and Technology}, 49(3):412--439, 2006.

\bibitem{Macchi_2_RMP}
A.~Macchi, M.~Borghesi, and M.~Passoni.
\newblock{} Ion acceleration by superintense laser-plasma interaction.
\newblock{} {\em Rev. Mod. Phys.}, 85:751--793, May 2013.

\bibitem{Daido}
H.~Daido, M.~Nishiuchi, and A.~S. Pirozhkov.
\newblock{} Review of laser-driven ion sources and their applications.
\newblock{} {\em Reports on Progress in Physics}, 75(5):056401, 2012.

\bibitem{Fritzler}
S.~Fritzler, V.~Malka, G.~Grillon, J.~P. Rousseau, F.~Burgy, E.~Lefebvre, E.~d'Humi\'eres, P.~McKenna, and K.~W.~D. Ledingham.
\newblock{} Proton beams generated with high-intensity lasers: Applications to medical isotope production.
\newblock{} {\em Applied Physics Letters}, 83(15):3039--3041, 2003.

\bibitem{Ceccotti}
T.~Ceccotti, A.~L\'evy, H.~Popescu, F.~R\'eau, P.~D'Oliveira, P.~Monot, J.~P. Geindre, E.~Lefebvre, and Ph. Martin.
\newblock{} Proton acceleration with high-intensity ultrahigh-contrast laser pulses.
\newblock{} {\em Phys. Rev. Lett.}, 99:185002, Oct 2007.

\bibitem{Zeil}
K.~Zeil, S.~D. Kraft, S.~Bock, M.~Bussmann, T.~E. Cowan, T.~Kluge, J.~Metzkes, T.~Richter, R.~Sauerbrey, and U.~Schramm.
\newblock{} The scaling of proton energies in ultrashort pulse laser plasma acceleration.
\newblock{} {\em New Journal of Physics}, 12(4):045015, 2010.

\bibitem{Spencer}
I.~Spencer, K.~W.~D. Ledingham, P.~McKenna, T.~McCanny, R.~P. Singhal, P.~S. Foster, D.~Neely, A.~J. Langley, E.~J. Divall, C.~J. Hooker, R.~J. Clarke, P.~A. Norreys, E.~L. Clark, K.~Krushelnick, and J.~R. Davies.
\newblock{} Experimental study of proton emission from 60 fs, 200 {mJ} high-repetition-rate tabletop-laser pulses interacting with solid targets.
\newblock{} {\em Phys. Rev. E}, 67:046402, Apr 2003.

\bibitem{Neely}
D.~Neely, P.~Foster, A.~Robinson, F.~Lindau, O.~Lundh, A.~Persson, C.-G. Wahlstr\"om, and P.~McKenna.
\newblock{} Enhanced proton beams from ultrathin targets driven by high contrast laser pulses.
\newblock{} {\em Applied Physics Letters}, 89(2), 2006.

\bibitem{Yogo}
A.~Yogo, H.~Daido, S.~V. Bulanov, K.~Nemoto, Y.~Oishi, T.~Nayuki, T.~Fujii, K.~Ogura, S.~Orimo, A.~Sagisaka, J.-L. Ma, T.~Zh. Esirkepov, M.~Mori, M.~Nishiuchi, A.~S. Pirozhkov, S.~Nakamura, A.~Noda, H.~Nagatomo, T.~Kimura, and T.~Tajima.
\newblock{} Laser ion acceleration via control of the near-critical density target.
\newblock{} {\em Phys. Rev. E}, 77:016401, Jan 2008.

\bibitem{Flacco}
A.~Flacco, F.~Sylla, M.~Veltcheva, M.~Carri\'e, R.~Nuter, E.~Lefebvre, D.~Batani, and V.~Malka.
\newblock{} Dependence on pulse duration and foil thickness in high-contrast-laser proton acceleration.
\newblock{} {\em Phys. Rev. E}, 81:036405, Mar 2010.

\bibitem{ALaDyn_GPL_20171123}
S.~Sinigardi, P.~Londrillo, A.~Marocchino, and A.~Sgattoni.
\newblock{} ALaDyn PIC Code.
\newblock{} {\tt doi:10.5281/zenodo.1065413}, 2017.

\end{thebibliography}


\end{document}